\documentclass[journal=cmatex,manuscript=article, layout=onecolumn]{achemso}

\usepackage{chemformula} % Formula subscripts using \ch{}
\usepackage[T1]{fontenc} % Use modern font encodings
\usepackage[utf8]{inputenc}
\newcommand{\degree}{$^{\circ}$}
\author{Sarah Hoffmann-Urlaub}
\affiliation[Georg-August University of G{\"o}ttingen]
{Institute for Materials Physics, Friedrich-Hund Platz 1, Georg-August University of G{\"o}ttingen, Germany}
\email{shoffma3@gwdg.de} 
\author{Ulrich Ross}
\affiliation[Georg-August University of G{\"o}ttingen]
{Institute for Materials Physics, Friedrich-Hund Platz 1, Georg-August University of G{\"o}ttingen, Germany}
				\author{J{\"o}rg Hoffmann}
				\affiliation[Georg-August University of G{\"o}ttingen]
{Institute for Materials Physics, Friedrich-Hund Platz 1, Georg-August University of G{\"o}ttingen, Germany}
			  \author{Alexandr Belenchuk}
				\affiliation[IEEN]
{Institute of Electronic Engineering and Nanotechnologies, Academiei str., 3/3, MD-2028, Chisinau, Republic of Moldova}
				\author{Oleg Shapoval}
\affiliation[IEEN]
{Institute of Electronic Engineering and Nanotechnologies, Academiei str., 3/3, MD-2028, Chisinau, Republic of Moldova}
				\author{Vladimir Roddatis}
				\affiliation[Georg-August University of G{\"o}ttingen]
{Institute for Materials Physics, Friedrich-Hund Platz 1, Georg-August University of G{\"o}ttingen, Germany}
				\author{Qian Ma}	
				\affiliation[Georg-August University of G{\"o}ttingen]
{Institute for Materials Physics, Friedrich-Hund Platz 1, Georg-August University of G{\"o}ttingen, Germany}
				\author{Birte Kressdorf}
				\affiliation[Georg-August University of G{\"o}ttingen]
{Institute for Materials Physics, Friedrich-Hund Platz 1, Georg-August University of G{\"o}ttingen, Germany}
				\author{Vasily Moshneaga}
				\affiliation[Georg-August University of G{\"o}ttingen]
{I. Institute of Physics, Friedrich-Hund Platz 1, Georg-August University of G{\"o}ttingen, Germany}
				\author{Christian Jooss}
				\affiliation[Georg-August University of G{\"o}ttingen]
{Institute for Materials Physics, Friedrich-Hund Platz 1, Georg-August University of G{\"o}ttingen, Germany}

\title[Tailoring c-axis orientation in epitaxial Ruddlesden-Popper Pr$_{0.5}$Ca$_{1.5}$MnO$_4$ films]
  {Tailoring c-axis orientation in epitaxial Ruddlesden-Popper Pr$_{0.5}$Ca$_{1.5}$MnO$_4$ films}
\abbreviations{RP, Ruddlesden-Popper; PV,
perovskite; RS, rock-salt; TEM, transmission
electron microscopy; 
EELS, electron energy loss spectroscopy; STEM, scanning
transmission electron microscopy; XRD, X-ray diffraction; HAADF, high angle annular dark field;
ADF, annular dark field}

\begin{document}

%\begin{tocentry}

		%\includegraphics[width=8.25cm]{Figures/TOC.png}%Höhe 4.45

%\end{tocentry}

\begin{abstract}
Interest for layered Ruddlesden-Popper strongly correlated manganites of \linebreak Pr$_{0.5}$Ca$_{1.5}$MnO$_4$ as well as to their thin film polymorphs is motivated by the high temperature of charge orbital ordering above room temperature. We report on the tailoring of the c-axis orientation in epitaxial RP-PCMO films grown on SrTiO$_3$ (STO) substrates with different orientations as well as the use of CaMnO$_3$ (CMO) buffer layers. Films on STO(110) reveal in-plane alignment of the c-axis lying along to the [100] direction. On STO(100), two possible directions of the in-plane c-axis lead to a mosaic like, quasi two-dimensional nanostructure, consisting of RP, rock-salt and perovskite building blocks. With the use of a CMO buffer layer, RP-PCMO epitaxial films with c-axis out-of-plane were realized. Different physical vapor deposition techniques, i.e. ion beam sputtering (IBS), pulsed laser deposition (PLD) as well as metalorganic aerosol deposition (MAD) are applied in order to distinguish between the effect of growth conditions and intrinsic epitaxial properties. For all deposition techniques, despite their very different growth conditions, the surface morphology, crystal structure and orientation of the thin films reveal a high level of similarity as verified by X-ray diffraction, scanning and high resolution transmission electron microscopy. We found that for different epitaxial relations the stress in the films can be relaxed by means of a modified interface chemistry. The charge ordering in the films estimated by resistivity measurements occurs at a temperature close to that expected in bulk material. 

\end{abstract}

\section{Introduction}
Charge ordered 3D perovskite manganites such as Pr$_{1-x}$Ca$_x$MnO$_3$ (PCMO) have been investigated intensively due to their unique electrical and magnetic properties, e.g. colossal magneto\cite{rao1998colossal} and electroresistance\cite{asamitsu1997current,westhauser2006comparative}, photoinduced phase transitions\cite{fiebig1998visualization,beaud2014time} and thermoelectricity\cite{cong2004high}. Recently, layered Ruddlesden-Popper (RP) perovskites have attracted considerable attention as they display novel physical properties directly related to their 2D structure, such as high activity and stability during oxygen evolution reaction\cite{ebrahimizadeh2016oxygen} and a charge ordering temperature above room temperature\cite{ibarra2003charge}, making them a model system for energy and catalytic applications. \\
The RP-phases are described by the general formula A$_{n+1}$B$_n$X$_{3n+1}$, where \textit{n} is the order of the RP-phase denoting the number of 2D perovskite layers that are enclosed by rock-salt (A-O) layers along the stacking direction. The $n=\infty$ corresponds to the 3D perovskite, e.g. PCMO. In contrast to bulk perovskites, epitaxial thin films can offer modified microstructure related properties due to growth-induced defects and lattice misfit strain to the substrate. The film microstructure can thus be tailored by growth on substrates with different orientations\cite{chu2007domain,lacotte2014growth} and changing lattice constants\cite{majumdar2012evolution}. In general, the properties of transition metal oxides depend strongly on point defects such as oxygen vacancies\cite{raabe2012situ,seong2009effect}, which in turn are related to the preparation method. Moreover, electronic and magnetic ordering at the nanoscale are affected by lattice distortions\cite{cao2011strain,herklotz2014strain} mediated by cooperative tilting and rotation of the MnO$_6$ octahedra, which yields a change of charge carrier mobility\cite{huang2016effect} and electrical resistance\cite{zheng2007converse} as well as the charge ordering temperature\cite{yang2006enhancing}.\\
Epitaxial thin films based on nickelates (La$_{n+1}$Ni$_n$O$_{3n+1}$)\cite{wu2013epitaxial,burriel2007enhanced}, titanates (Sr$_{n+1}$Ti$_n$O$_{3n+1}$)\cite{shibuya2008sr,jungbauer2014atomic,gutmann2006oriented,kamba2003composition}, strontium oxides (La$_{n+1}$Sr$_n$O$_{3n+1}$)\cite{takamura2006thickness,palgrave2012artificial} and manganites (La$_{n-nx}$Ca$_{1+nx}$Mn$_n$O$_{3n+1}$)\cite{asano1997magnetotransport} or \linebreak (La$_2$Sr$_2$Mn$_3$O$_{10}$)\cite{palgrave2012artificial} have been grown by co-deposition up to $n\,=\,3$. The limited thermodynamic stability of RP phases with n$\geq$4 requires an ''atomic-layer-epitaxy'' growth\cite{lei2017constructing,lyzwa2018situ} with sequential deposition of the perovskite and rock-salt monolayers therefore creating an enforced generation of the stacking sequence along $\left[001\right]$-direction. \\
Here we are focusing on thin films of A-site half-doped ($Pr/Ca=1$) first order ($n=1$) system  Pr$_{0.5}$Ca$_{1.5}$MnO$_4$ (RP-PCMO) prepared by physical vapor technique based co-deposition. Experiments solely on RP-PCMO nanoparticles have been reported so far\cite{mierwaldt2017environmental}. However, energy conversion and storage applications often require the availability of the RP material as thin films, where thin film epitaxy is highly desirable for the preparation of well-defined junctions. Considering the desire for photovoltaic thin film applications (e.g. \cite{luo2006rectifying}) with electrically conducting Nb-doped SrTiO$_3$ substrates, we focus here on the growth on STO substrates with (001) and (110) orientations in order to access multiple strain states in RP-PCMO/STO films. In addition, we have studied the growth of RP-PCMO on a CaMnO$_3$ (CMO) buffer layer so as to release the lattice misfit strain. To distinguish between the preparation- and substrate-related effects in view of their impact on the growth mechanisms and characteristics, the films were grown by means of fundamentally different deposition techniques, i.e. ion beam sputtering (IBS), pulsed laser deposition (PLD) and metalorganic aerosol deposition (MAD). 

\section{Results and discussion}
For each deposition method, a suitable parameter set was identified to grow phase pure, homogeneous and smooth thin films in the desired stoichiometry and crystal structure (see Fig. \ref{fig:XRD_100_All_large}). Thus, on the same type of substrate very similar results are obtained for all three methods with respect to the film morphology (see Fig. \ref{fig:SEM_AFM}) and structure. Hence, not all results from samples grown by all used techniques are shown, but only the most-suited films are presented. \\
The lattice constant of the cubic STO substrate is a$_{STO}$\,=\,3.906\,\AA, while for the RP phase in the space group $Fmm2$ the values of a$_{RP}$\,=\,5.365\,\AA, b$_{RP}$\,=\,5.354\,\AA\, and c$_{RP}$\,=\,11.840\,\AA\, are found in bulk materials\cite{ebrahimizadeh2016oxygen,chi2007effect}. This implies that there are no lowest common multiples for any combinations of axes that would enable a reasonably good matching of both materials to enable a commensurate epitaxy. Between the c-axis of the RP phase (c$_{RP}$\,=\,11.840\,\AA) and tripled length of the a$_{STO}$\,=\,3.906\,\AA\, a lattice mismatch of only 1\% would be expected, whereas along the a- and b- directions the lattice mismatch between the RP-PCMO film and STO substrate is significantly larger. Parallel to [010]/[001] of the substrate the mismatch is >30\% and 3\% parallel to [110], hence the growth of a completely epitaxial film is unfavorable and the growth of small grains/domains is be expected.\\
Next, we consider the film growth on STNO(110), since in this case a more suitable substrate lattice parameter of 5.525\,\AA\, is provided in one direction. The anisotropic surface structure supports a preferred growth direction of the film. Due to the asymmetry of the lattice parameters in the substrate, an in-plane growth of the c-axis is expected. The mismatch of the a- or b-axis of the RP-phase parallel to the [110]-direction of the substrate is -3\% and 1\% for the c-axis parallel to the [001]-direction (compare Fig. \ref{fig:S_Lattice_matching}b). Due to the evenly matched lattice parameters of the a- and b-axis, the formation of both [100] and [010] crystallites/domains is equiprobable.

\begin{figure*}[!ht]
\includegraphics*[width=\textwidth]{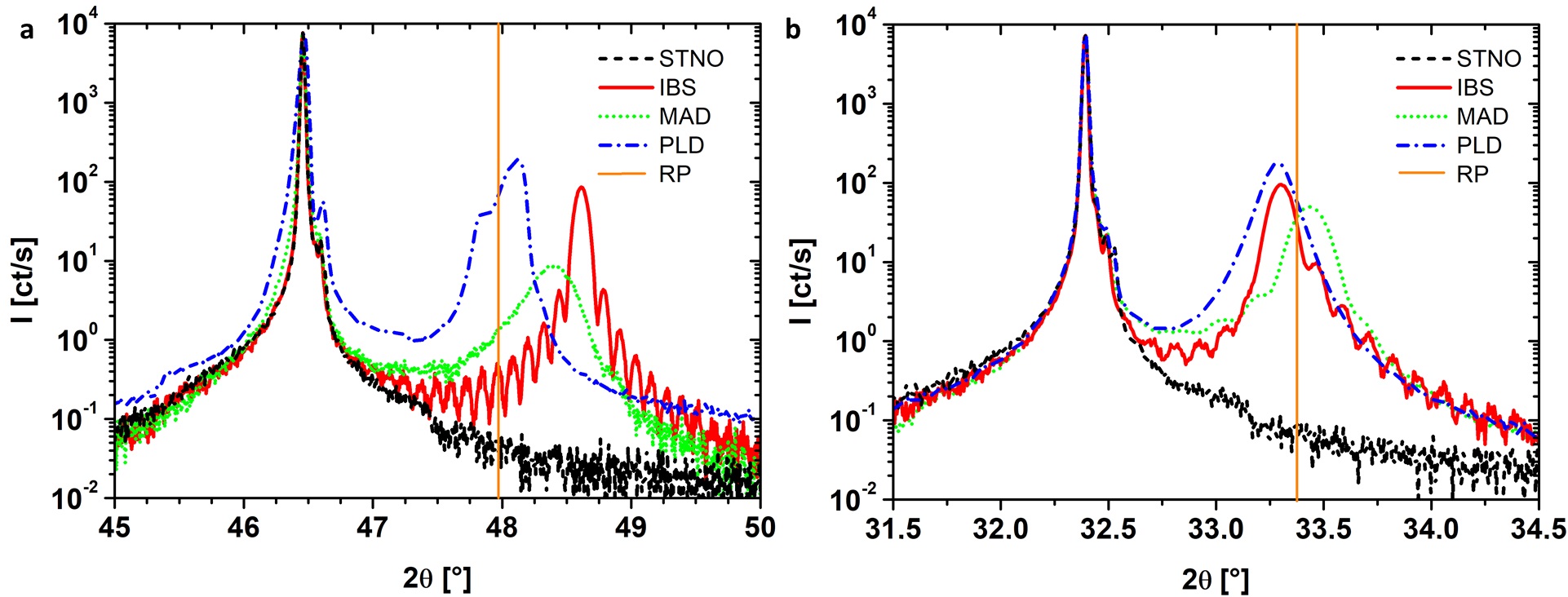} 

\caption{XRD patterns of thin films grown by different techniques on: a) STO(001) substrates with a detailed view on the (220) reflection of the films and b) on STNO(110) substrates with an enlarged view on the (020) reflection of the film. The black curves show the scans of the bare substrates and the orange lines represent the nominal RP peak positions. All intensities are normalized to the substrate peak.}%MAD 35nm IBS 100 nm  PLD 130nm
\label{fig:XRD}
\end{figure*}
For both cases the crystal structure and orientation of the films are examined by means of XRD (see Fig. \ref{fig:XRD}). Note that, as the RP [100] and [010] directions cannot be distinguished, only one of them is indexed for reasons of clarity, while for the reference the quadratic mean is used. \\
As shown by the XRD patterns in Fig. \ref{fig:XRD}, the peak positions of the IBS and MAD films grown in (220) orientation noticeably differ from the nominal RP bulk value of 47.97\degree, whereas in (010)-orientation all peaks are in good agreement with the expected value of 33.376\degree\, (see Tab. \ref{tab:XRD}), indicating less strained films. For the (220) orientation only tensile stress (negative strain values) is observed for all deposition techniques, while the variations in the absolute strain values are most likely originating from the differences in the film thicknesses and the deposition temperatures. In contrast, the strain (if any) in the (020) grown films is flipped to compressive for IBS and PLD, while the absolute values are reduced compared to the previous growth direction. \\
The observed crack microstructure (see Fig. \ref{fig:SEM_AFM}a) appears to be responsible for the release of the in plane tensile stress. Despite the partial reduction of this strain via crack formation and the presence of other stress relaxation channels such as point or planar defects, the films still possess a large out-of-plane strain. Although the nominal mismatch of the RP-phase to the lattice is the same, the films on STNO(110) with much smaller out-of-plane stress and compressive strain of the order of 0.2-0.3\% are consequently crack-free. The out-of-plane strain values for all films and both substrate orientations are listed in Table \ref{tab:XRD}. 

\begin{table}[!h]  
		\centering
  \caption{Analysis of the X-ray scans in Fig. \ref{fig:XRD} for the RP-PCMO films on STO (100) (left column) and STNO (110) (right column). For the calculation of the strain $\epsilon=\frac{d-d_0}{d_0}$ is used, where $d_0$ is the nominal lattice parameter and $d$ of the film.}
  \begin{tabular}[htbp]{c|cc|cc}
    \hline
    Deposition &  Position                & Strain [\%] & Position               & Strain [\%]\\
		 Technique &  (220) reflex [\degree]  &             &  (020) reflex [\degree]&            \\
\hline
    PLD    & 48.14 & -0.32&33.28&0.27\\
		IBS    & 48.61 & -1.24&33.30&0.22\\
    MAD    & 48.39 & -0.82&33.43&-0.16\\
    \hline
  \end{tabular}
  \label{tab:XRD}
\end{table}

Furthermore, the occurrence of Laue fringes, most pronounced for the IBS and MAD films, is an indication for the uniformity of the lattice parameter and, hence, imply that these films are homogeneously strained. From the analysis of the fringes (e.g. \cite{ying2009rigorous}) the estimated size of crystalline domains is ~12-15\,nm for the 35-100\,nm thick films in both orientations. This value denotes the lower limit of the size of the coherently oriented but slightly tilted or twisted crystal domains. \\
To understand the interplay of crystal orientation, strain and domain and/or crack formation on an atomic scale level we use high-resolution (scanning) transmission electron microscopy HR(S)TEM to analyze the microstructure of RP-PCMO films grown on both substrate orientations.\newpage

\begin{figure*}[!h]
\includegraphics*[width=\textwidth]{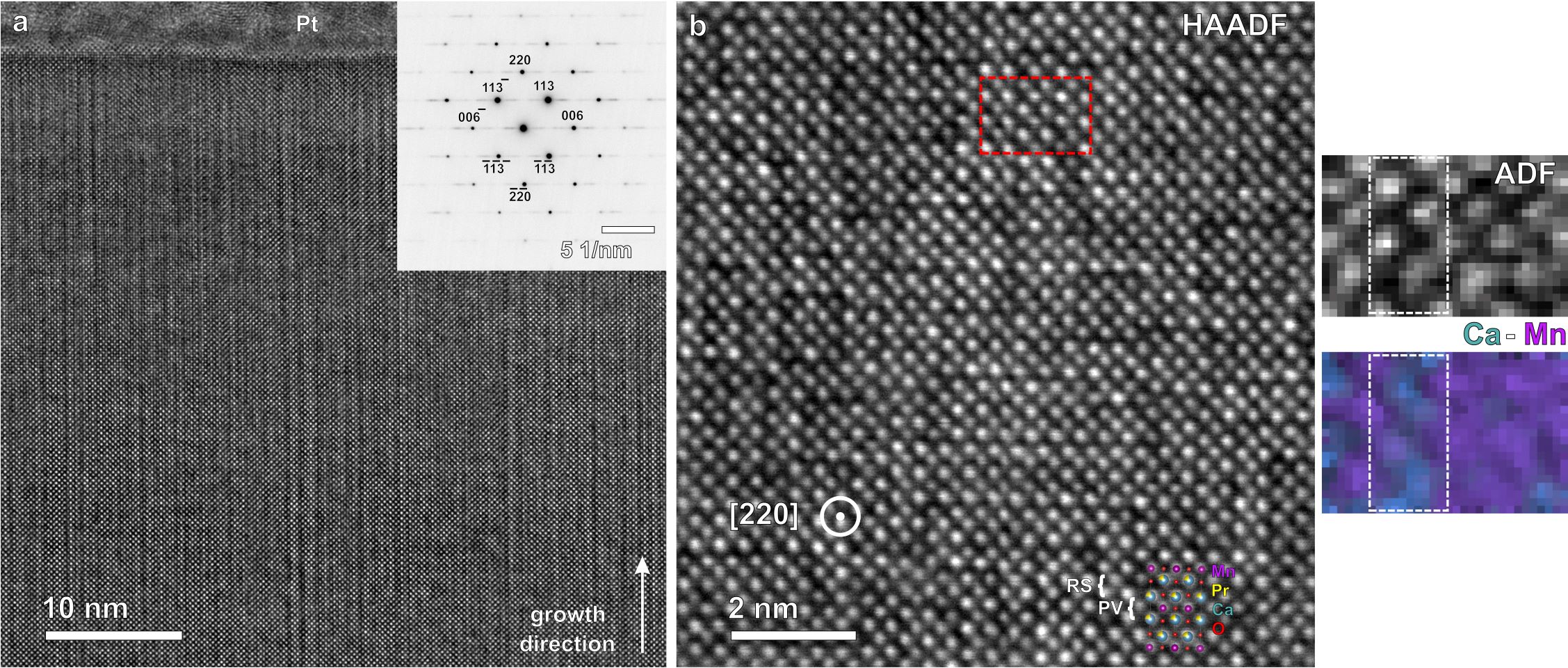}
\caption{IBS thin film grown on STO(001). a) HRTEM cross-section of the film, capped by a Pt protection layer. Inset: Area-averaged nanodiffraction. b) HAADF-STEM plane view of the film shown in a) with drawn in positions of the atomic columns, forming a RP-unit cell. c) Top: ADF-STEM image of the region marked in b), below: STEM-EELS map of Ca-Mn. The white dashed boxes mark the position of the rock-salt layer.}
\label{fig:TEM100}
\end{figure*}
\begin{figure*}[!h]
		\centering
			\includegraphics*[width=1.00\textwidth]{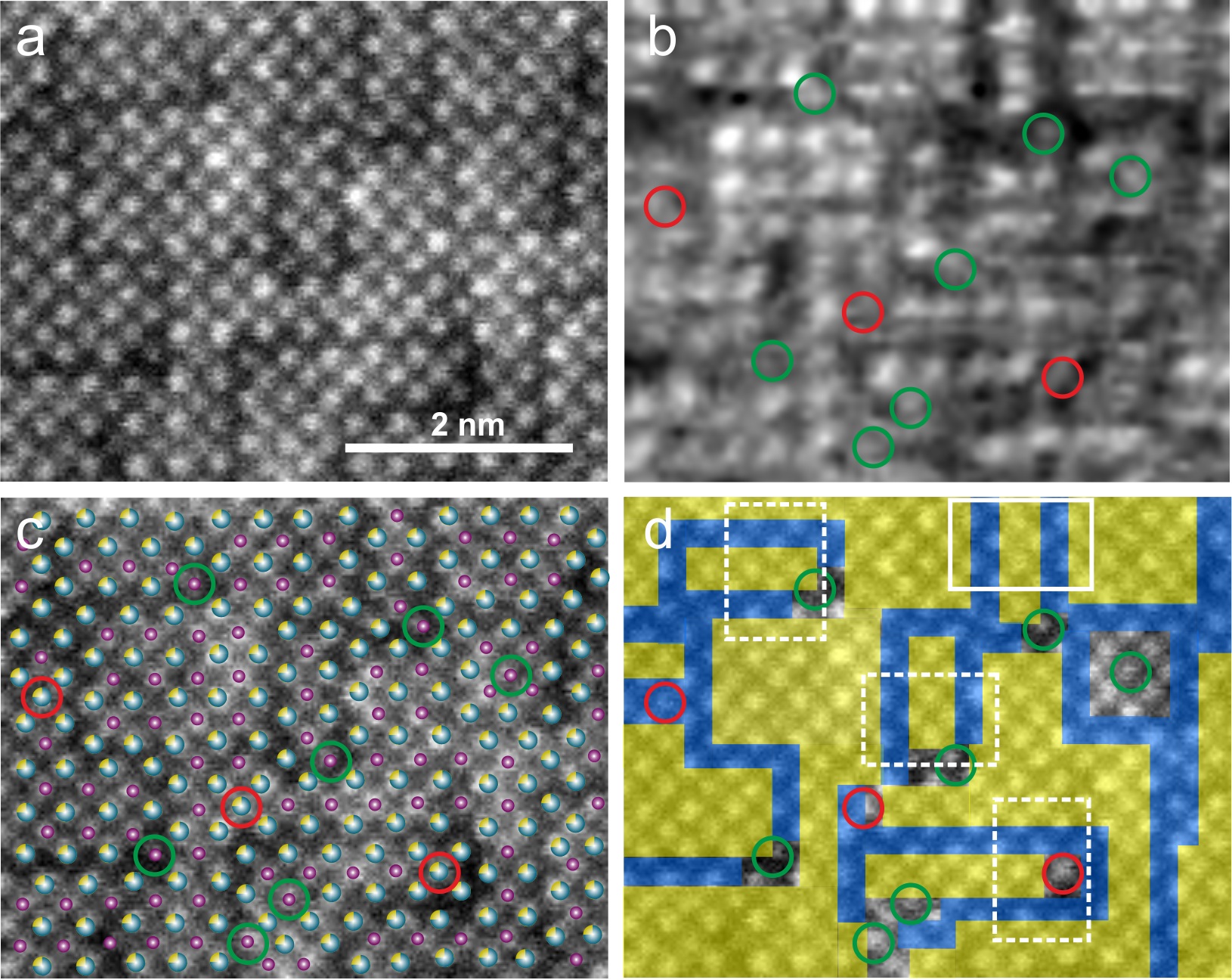}
		\caption{Analysis of nanoscopic mosaic type ordering of the two structural units (rock-salt and perovskite) in a RP-PCMO film grown on STO(001). All images show the same area. (a) STEM-ADF image. (b) EELS map of the Mn L-edge at 640\,eV. In a qualitative manner, green circles mark Mn placed on Pr or Ca sites, while red circles mark the opposite situation of Pr or Ca placed on Mn sites. (c) STEM-ADF image with atomic species drawn in, based upon the correlation of (a) and (b). Mn atoms are drawn in purple and Pr/Ca in green and yellow. (d) Topology of rock-salt layers (blue) and perovskite blocks (yellow). The white boxes mark the positions of RP-units,dashed lines circumvent incomplete cells and the solid line a complete unit cell.}
		\label{fig:Mosaic}
	\end{figure*}

In Fig. \ref{fig:TEM100}a) we show a cross-section HRTEM image of a film on STO(001) cut parallel to the [001] direction of the substrate. The diffraction information in the inset confirms the XRD results of (220) out-of-plane orientation (see Fig. \ref{fig:XRD}a) and shows that the c-axis [00\textit{l}] of the RP film lies in-plane. However, an irregular layering is observed perpendicular to the growth direction. Hence, streaks occur in the diffraction pattern as a result of continuous variations of the lattice plane distances. To understand this in more detail, a planar view on the structure was examined using a high-angle annular dark field (HAADF) STEM as shown in Fig. \ref{fig:TEM100}b). The in-plane nanostructure is characterized by the arrangement of the two constituent structural units, i.e. perovskite (PV, Mn-rich) and rock-salt (RS, CaO) layers, respectively, which can be identified from the annular dark field (ADF) image and electron energy loss spectroscopy (EELS) mapping in Fig. \ref{fig:TEM100}c). The nanostructure consists of a mosaic-like RP domain pattern with two different orientations of the RP c-axis, where individual domains of uniform RP stacking / RP c-axis orientation can extend over up to five unit cells. However, it is observed that stacking faults locally form perovskite blocks extended over multiple unit cells. In this configuration the direction of the c-axes is two-fold degenerated with alignment parallel to either the [100] or the [010] axis of the substrate. The 90\degree changes in c-orientation are forming a network of rock-salt layers throughout large parts of the film (see Fig. 2a). As described above, the in-plane strain on the RP-unit cell caused by the lattice mismatch is anisotropic\footnote{Compressive along the c-axis and tensile in the perpendicular directions.}(compare Fig.\ref{fig:S_Lattice_matching}), hence the formation of a mosaic pattern with alternating alignment of the c-axis prevents the buildup of large crystalline domains and thus reduces the misfit strain. As we did not find any sign of phase intermixing/segregation (this is a common way to relax the stress in this materials system\cite{ifland2015strain}) throughout the whole thickness of the film, it can be denoted as a homogeneous mosaic-type RP phase. The observation of this type of quite uniform network of RP layers leads to the conclusion that this ordered structure grows by means of self-organization with preservation of the mean stoichiometry on the one hand and accommodation of in-plane (biaxial) misfit stress on the other hand. \\
To investigate the mosaic-like arrangement of the structural units in more detail, we show in Fig. \ref{fig:Mosaic} HRSTEM imaging (a) complemented by spectroscopy (b) to assign the different chemical species on the A-site (Pr/Ca) and B-site (Mn) to their local positions in the crystal structure (c). 
\newpage
After identifying the two structural units of rock-salt and perovskite it is apparent that the rock-salt layer network is interrupted by partial dislocations.\\
It is notable that the (incomplete) unit cells are separated from each other and that for two closely spaced cells a 90\degree-rotated orientation is preferred. Both topological aspects directly correlate with to the lattice mismatch between the film and the substrate, since the strain on two isolated cells is smaller compared to that of two merged cells that would span over a twice as large substrate area. In addition, the alternating arrangement prevents the accumulation of compressive or tensile stress in one direction.\\
The corresponding films show a network of cracks (see Fig. \ref{fig:SEM_AFM}a), aligned rectangular with respect to each other as well as to the <100>- directions of the STO crystal. This microscopic effect can be directly understood from the nanoscale arrangement of the two structural units. As a matter of principle, the bond length across a rock-salt layer is larger compared to that of the perovskite layer\cite{battle1998chemistry}. We assume that the rock-salt bonding is weaker, as is well known from their tendency to accept interstitial oxygen\cite{nakamura2016determining}. Due to this, it is most probable that the cracks run along the rock-salt layers, and these cracks are even extended into the substrate. This indicates a strong adhesion between the substrate and the film as was documented for this material system earlier\cite{herpers2014competing}. \\

The defect and domain structure on STNO(110) substrates is quite different, hence the RP-PCMO growth is investigated by TEM analysis to gain information about the ordering of the structural units on a substrate with anisotropic surface lattice.
\begin{figure*}[!h]
\includegraphics*[width=\textwidth]{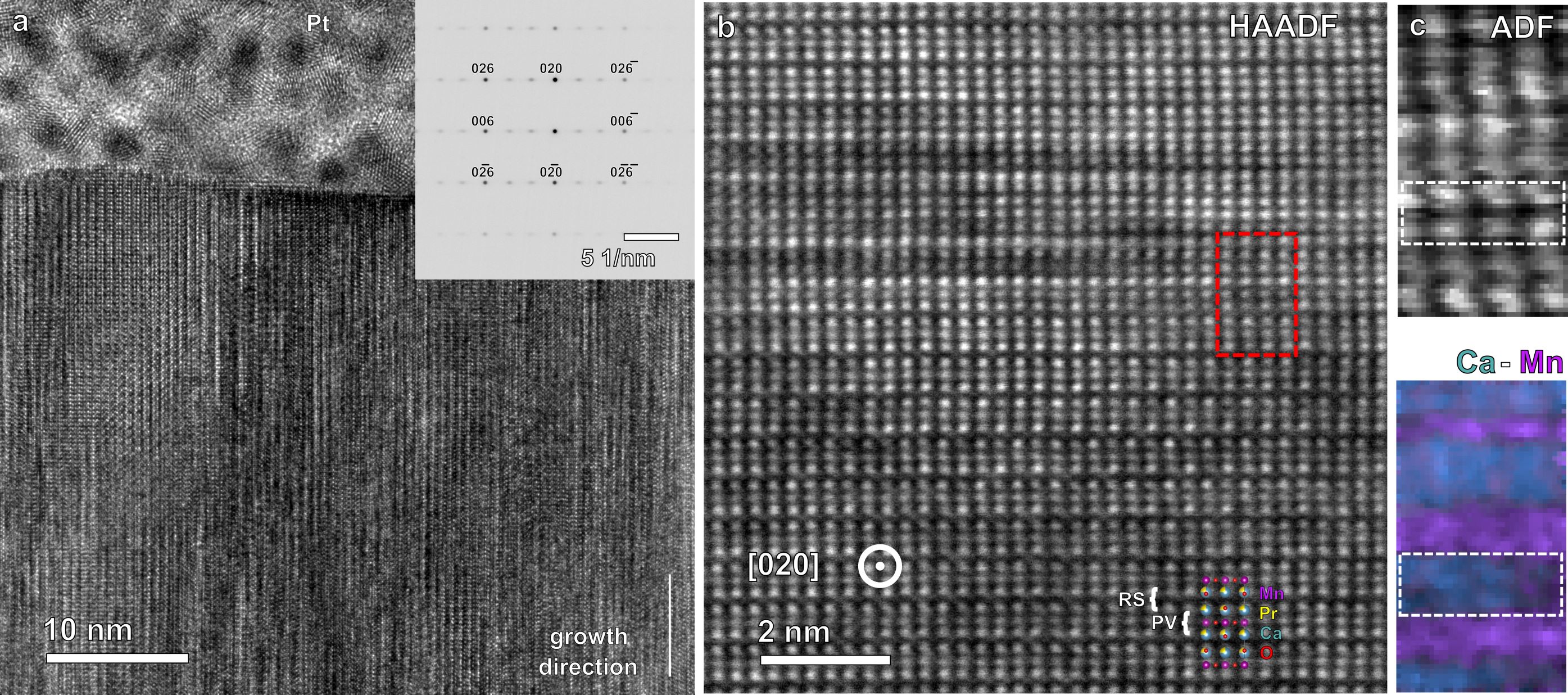}
\caption{IBS-grown thin film on a STNO(110) substrate. a) HRTEM cross-section of the film, capped by a Pt-layer. Inset: Area-averaged nanodiffraction. b) HAADF-STEM plane view of the film in a) with drawn in positions of the atoms, forming a RP-unit cell. c) Top: ADF-STEM image of the region marked in b), below: STEM-EELS map of Ca-Mn.} 
\label{fig:TEM110}
\end{figure*}

The PCMO-RP films in Fig. \ref{fig:TEM110}a) deposited on (110) oriented substrates show a high degree of ordering. The domains of RP phase are separated by stacking faults and possibly by grain boundaries with very small angles of less than 1\degree. They reveal spatial extensions of about 20\,nm, which is in good agreement to the coherence length of 10-15\,nm estimated by XRD analysis. In the planar view the alternate layering of perovskite and rock-salt layers is verified by EELS mapping revealing a stacking sequence of Ca-O rich as well as Mn-O rich layers (b). \\
As a fundamental result the films on STNO(110) are crack-free (see Fig. \ref{fig:SEM_AFM}b) and just as smooth as the films on (001) (d). The AFM analysis supports the findings in the SEM images, since the RMSE roughness accounts for 0.2\,nm, which is equal to the difference between two SrO/TiO$_2$ terrace steps of the unterminated substrate.\\
A different degree of strain in the mosaic RP phase in films on STO (100) and c-axis in-plane RP phase in films on STNO (110) is also supported by the results of x-ray diffraction. As revealed in table \ref{tab:XRD}, the out of plane a and b lattice parameter in the film with c-axis in-plane differ less from the nominal value than the strains in the (110) planes of the mosaic type film. The occurrence of the second peak for the film prepared by PLD in Fig. \ref{fig:XRD}a) may be an indication for a degenerated lattice parameter that may be attributed to the steric configuration of the RP unit cells in the mosaic pattern, while coherently only one peak is observed for the layered RP structure in the films on STNO (110).  \pagebreak 

\begin{figure*}[!ht]
\includegraphics*[width=0.48\textwidth]{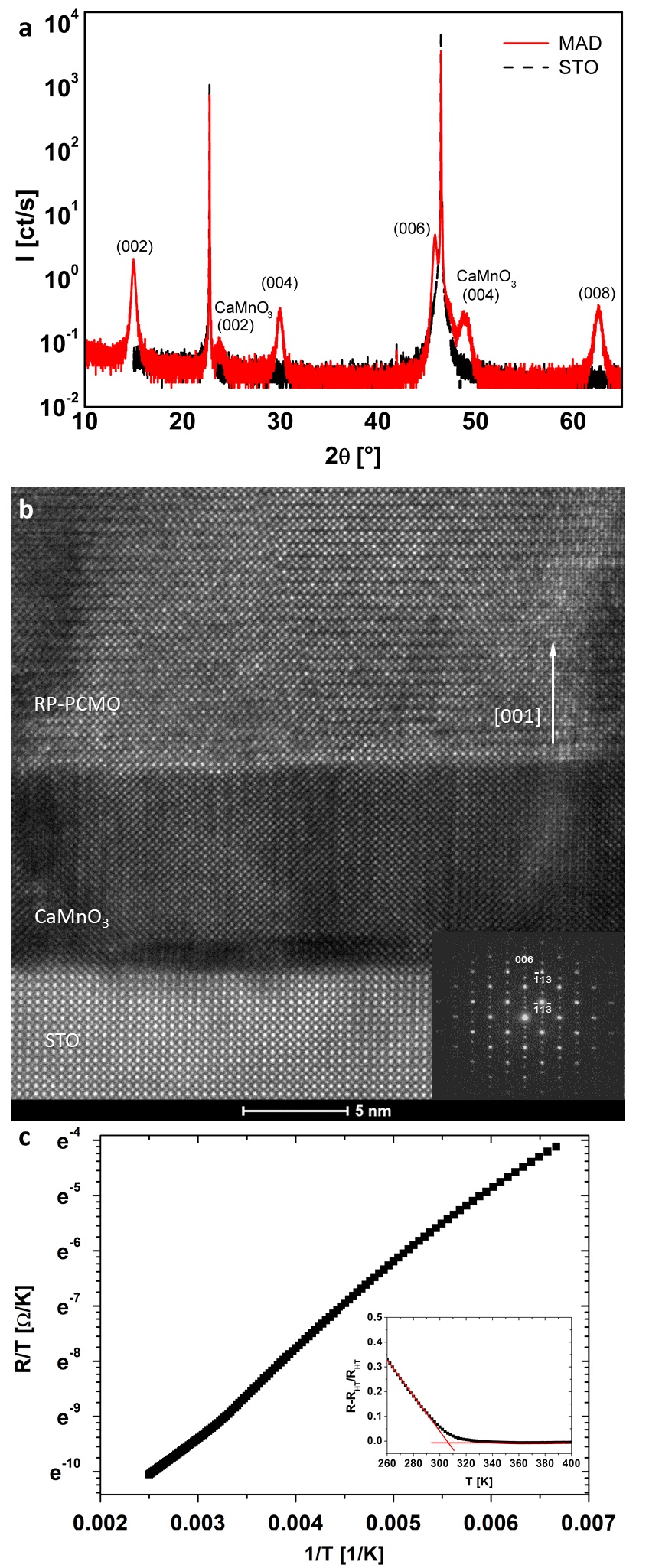}
\caption{MAD epitaxial RP-PCMO thin film grown on STO(100) with a CaMnO$_3$ buffer layer. a) Indexed x-ray diffractogramm. b) TEM cross-section with the corresponding electron diffraction pattern in the inset. c) In-plane temperature dependent resistivity measurement in Arrhenius representation. Inset: Normalized deviation of the measured resistivity \textit{R} from the linear fit of the high temperature region \textit{R}$_{HT}$ of the Arrhenius plot.}
\label{fig:C-axis}
\end{figure*}

\newpage

In Fig. \ref{fig:C-axis}a) we present the XRD pattern from a RP-PCMO film grown on a STO(001) substrate buffered by a 10\,nm thick CaMnO$_3$ (CMO) layer. The out-of-plane orientation of the c-axis is favored by introducing an [001] oriented orthorhombic CMO buffer layer\cite{poeppelmeier1982structure} with a pseudo-cubic lattice parameter of $d_{CMO}=3.72$\,\AA\ that is close to that of the RP-phase $d_{RP}=3.79$\,\AA. We assume that due to the notably reduced in-plane misfit stress of 1.7\% between RP-phase and buffer layer as well as the structural similarity of the lattice types an epitaxial growth is enabled. However, the [001] RP-PCMO growth was up to now only detected for MAD-grown films at an elevated growth temperature of 945\degree C. The length of the RP c-axis extracted from the XRD scans is 11.87\,\AA\,, which is in good agreement to the values given above. The size of domains of pure RP-phase was found to be up to a few tens of nanometers. They are separated by anti-phase boundaries induced by stacking faults at the interface to the CMO buffer layer (Fig. \ref{fig:C-axis}b)) that reveals a surface roughness smaller than one RP unit cell. \\ 
On basis of the accomplished growth of phase pure RP thin films, the charge order transition temperature is determined by measuring the in-plane resistivity as a function of temperature. This is exemplary depicted for the c-axis oriented MAD film (Fig. \ref{fig:C-axis}c). According to experimental data on bulk RP-PCMO the charge orbital ordering occurs at T$_{CO}\approx$ 320\,K (for instance \cite{mathieu2006bandwidth,handayani2015correlation}). For the studied RP-PCMO films on STO(110) and on STO (100) with the CMO buffer, a charge orbital ordering was also observed above room temperature as evidenced by the change of the slope in the resistivity curve in the Arrhenius representation and by the normalized deviation of the measured resistivity \textit{R} from the linear fit of the high temperature region \textit{R}$_{HT}$ of the Arrhenius plot. As a result the estimated value of 307\,K is found to be in excellent agreement with the data on bulk RP-PCMO.

\section{Conclusions}
In summary, epitaxial Pr$_{0.5}$Ca$_{1.5}$MnO$_4$ thin films grown by means of ion beam sputtering, pulsed laser and metalorganic aerosol deposition reveal similar properties in terms of surface morphology, crystal orientation, internal structure, strain states and defect density and types. Hence, no systematic impact of the deposition method on the growth characteristics was readily apparent. This study clearly confirms that by controlling the surface crystal symmetry or rather the lattice parameter of the underlying substrate the structural orientation and strain state of the RP-PCMO thin film grown on top can be adjusted. In detail, on STO(001) the growth of the RP-phase in [110] direction is observed, while the in-plane direction of the c-axis is degenerated leading to a mosaic like pattern of structural domains in the RP sub-unit cell range. Here, due to the mismatch to the cubic substrate lattice and the related tensile strain, a crack network aligned with respect to the underlying substrate is formed. On the STNO(110) substrate that exhibits an anisotropic surface lattice, the c-axis of the unit cells is arranged in parallel to the surface [100] direction. The films with an out of plane [020]/[200] orientation and are less strained and crack-free. To induce an out-of-plane c-axis orientation in MAD, a CaMnO$_3$ buffer layer was introduced. From the temperature dependent electrical resistance of RP-PCMO films, compelling evidence for a charge ordering temperature of T$_{CO}= 307$\,K was obtained, which is in excellent agreement with the bulk value of RP-PCMO.

\section{Experimental}
Dense ceramic targets of Pr$_{0.5}$Ca$_{1.5}$MnO$_4$ were prepared for PLD and IBS using standard ceramic synthesis methods. Appropriate ratios of Pr$_6$O$_{11}$ (CAS $12037-29-5$), CaCO$_3$ (CAS 471-34-1) and Mn$_2$O$_3$ (CAS 1317-34-6) powders (Sigma-Aldrich) were mixed and intimately ground using a planetary ball mill (Retsch). The powder was annealed twice for 12 h at 900\degree C and once for 12 h at 1100\degree C (Nabertherm, HTCT 01/16), with intermediate grindings. Pellets having a diameter of 2.5 cm were uniaxially cold pressed (Janssen HWP 20) at a pressure of 120\, bar for 20 minutes and sintered for 24 h at 1500\degree C. In between the process steps the homogeneity and phase purity of the powders and the final target is monitored by XRD. \\ 
For the MAD preparation the acetylacetonates of Pr, Ca and Mn were uses as precursors. The solution in organic solvent dimethylformaimde (DMFA) (Sigma, Aldrich) was prepared with experimentally determined (Pr+Ca)/Mn and Pr/Ca precursor ratios and concentration of 0.02 mole (calculated for Mn-precursor). Polished substrates of strontium titanate (STO) and the 0.5 wt \% doped Nb:SrTiO$_3$ (STNO) in (100) and (110) orientation were used (CrysTec).

\subsection{Deposition techniques}
PLD \cite{krebs2003pulsed} used a KrF excimer laser (COMPex 205F, Coherent) with a wavelength of 248\,nm, a pulse duration of 30\,ns and a repetition rate of 5\,Hz to fabricate thin films. The oxygen background pressure in the vacuum chamber is 0.17\,mbar, the laser fluence is kept constant at 1.36 J/cm$^2$ and the substrate temperature at 700\degree C.\\
IBS\cite{cuomo1989handbook} was carried out using a Kaufman source (2.5\,cm beam diameter, Ion Tech Inc.) with xenon as sputter gas. Beam current density and the beam voltage were fixed at (4\,mA$/$cm$^2$$/$1000\,V) while the partial pressure was $<$10$^{-4}$\,mbar and the substrate temperature 700\degree C.\\
For MAD\cite{moshnyaga1999preparation} the precursors were mixed in a suitable molar ratio and dissolved in an organic solvent (DMFA). The substrate was heated resistively by using a SiC heater up to 850\degree C and deposition was carried out at ambient air conditions using dried compressed air as spraying gas. \\
A suitable temperature window was chosen for all techniques to grow fully crystalline (Fig. \ref{fig:XRD_100_All_large}f) and smooth films (Fig. \ref{fig:SEM_AFM}). 
Furthermore, for the PLD and IBS as non-equilibrium techniques the oxygen partial pressure and energy density for photons (see Fig. \ref{fig:XRD_100_All_large}a)) and ions, respectively, were varied to promote growth in a certain crystal orientation by meeting the correct stoichiometry of the RP-phase (see Fig. \ref{fig:XRD_100_All_large}d)). For MAD as a close to equilibrium growth technique the precursor concentrations and solution feeding rates were adjusted for these purposes.

\subsection{Characterization methods}
X-ray diffraction (XRD, D8 Discovery, Bruker) was employed to analyze the crystallographic structure of targets and thin films. In the setup a G\"oebel mirror was installed to parallelize the Cu-K$\alpha$ radiation. The corresponding crystal structures were simulated by VESTA\cite{momma2008vesta}. A scanning electron microscope (SEM, Nova NanoSEM 650, FEI) was used to image the surfaces of the target and thin films. The system is furthermore equipped with a detector for energy dispersive X-ray spectroscopy (EDX, Oxford) to estimate their average atomic composition. With an atomic force microscope (AFM, MFP 3DSA, Asylum Research) the two dimensional surface morphology was investigated on the sub-nanoscale. Using a focused ion beam machine (FIB, Nova Nanolab 600, FEI) thin lamellae were extracted from the thin films for transmission electron microscopy (TEM, Titan 300\,keV, Gatan). At the same facility an Electron Energy Loss Spectrometer (EELS, Quantum ER/965P, Gatan) enables for spatially resolved composition analysis.

\begin{acknowledgement}

We We would like to thank Bent van Wingerden for his help with electron microscopy and X-ray diffraction, Benjamin Hauschild for PLD, Stephan Melles for IBS and Tobias Meyer and Michael Seibt for helpful discussions.\\
Funding by the Deutsche Forschungsgemeinschaft (DFG) via Grant No. 217133147/SFB1073 projects A02, B02 and Z02 and the use of equipment in the ''Collaborative Laboratory and User Facility for Electron Microscopy'' (CLUE) www.clue.physik.uni-goettingen.de are gratefully acknowledged.
\end{acknowledgement}

\bibliography{literature_list_IMP}
\begin{suppinfo}

 \setcounter{figure}{0}
%Brief descriptions in nonsentence format listing the contents of the files supplied as Supporting Information.
\renewcommand{\thefigure}{S\arabic{figure}}

\begin{figure}[!h]
		\centering
			\includegraphics[width=0.75\textwidth]{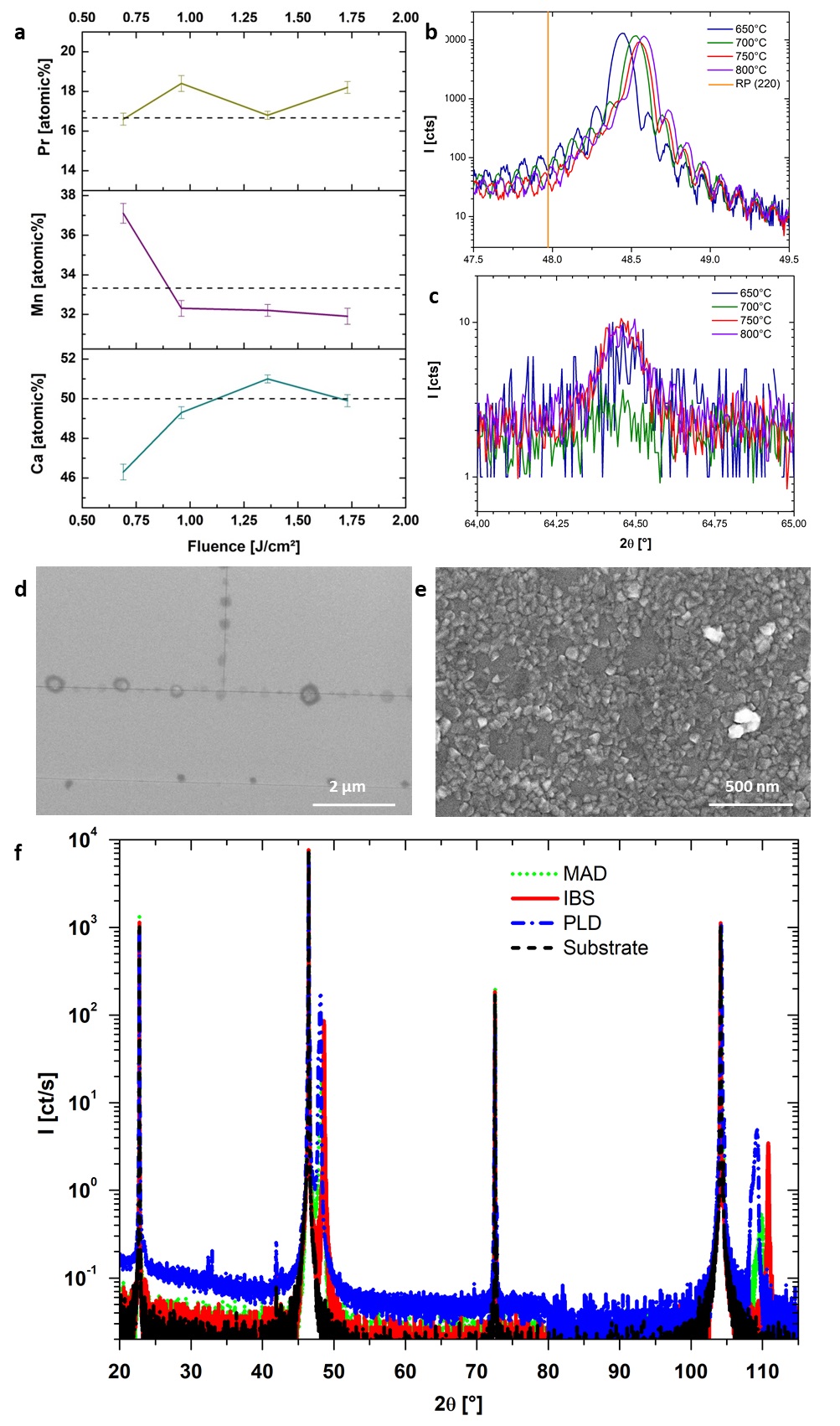}
			\caption{Optimization of thin films by estimating adequate growth parameters. (a) EDX analysis of 500\,nm thick films deposited at different laser fluences by PLD. (b) XRD scans of IBS films on STO(100). (c) Signal of a foreign phase is minimized at 700\degree C. (d,e) SEM images of films in (b) grown at 650\degree C and 800\degree C, respectively. (f) XRD scans for a larger 2$\Theta$ range for all techniques. The thin films are phase pure up to thicknesses of 100\, nm for IBS and PLD and 50\,nm for MAD and are grown on STO(100).}
			\label{fig:XRD_100_All_large}
\end{figure}

\begin{figure}[!h]
\includegraphics[width=\textwidth]{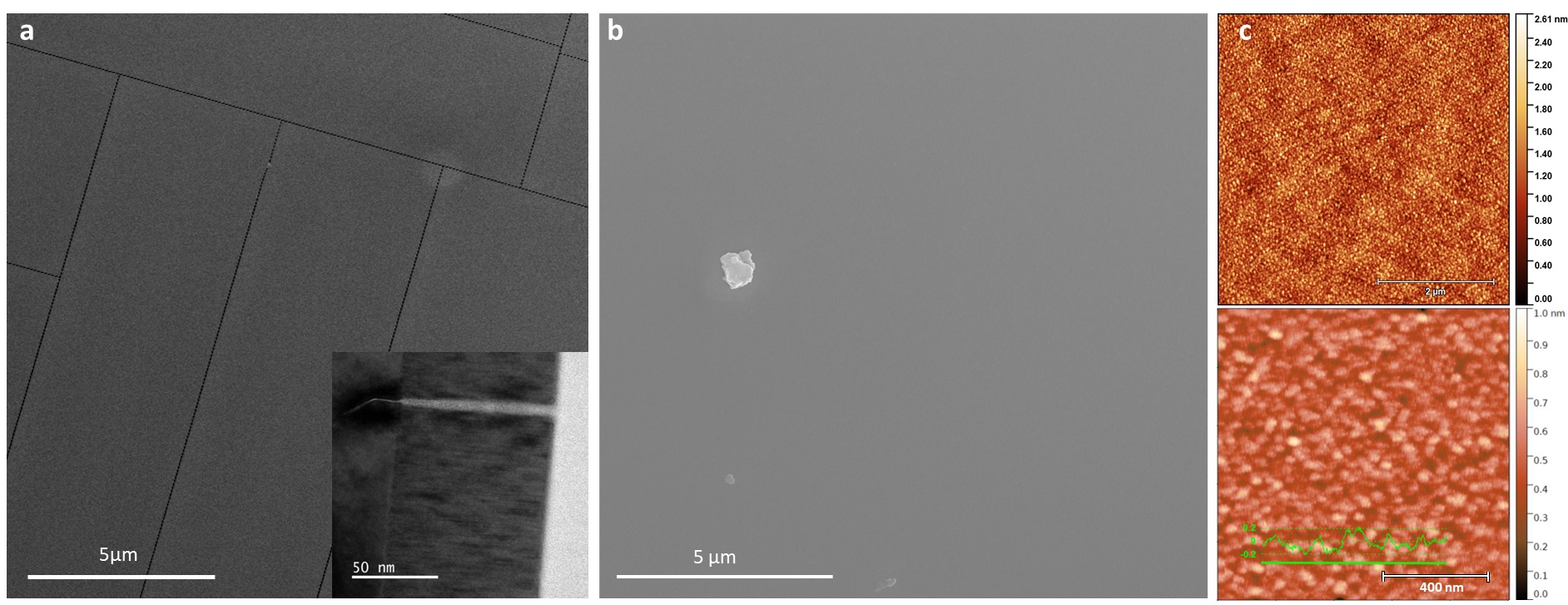}
\caption{Surface characterization of optimized films. Thin films a) on STO(001) an b) on STO(110). c) AFM images of IBS films. Top: on STO(001) and bottom: on STO(110) along with a line profile. The inset in a) shows a TEM image of a crack in a 75\,nm thick IBS deposited film that is extended into substrate. All films were grown at 700\degree C.}
\label{fig:SEM_AFM}
\end{figure}

	\begin{figure}[!h]
		\centering
			\includegraphics[width=\textwidth]{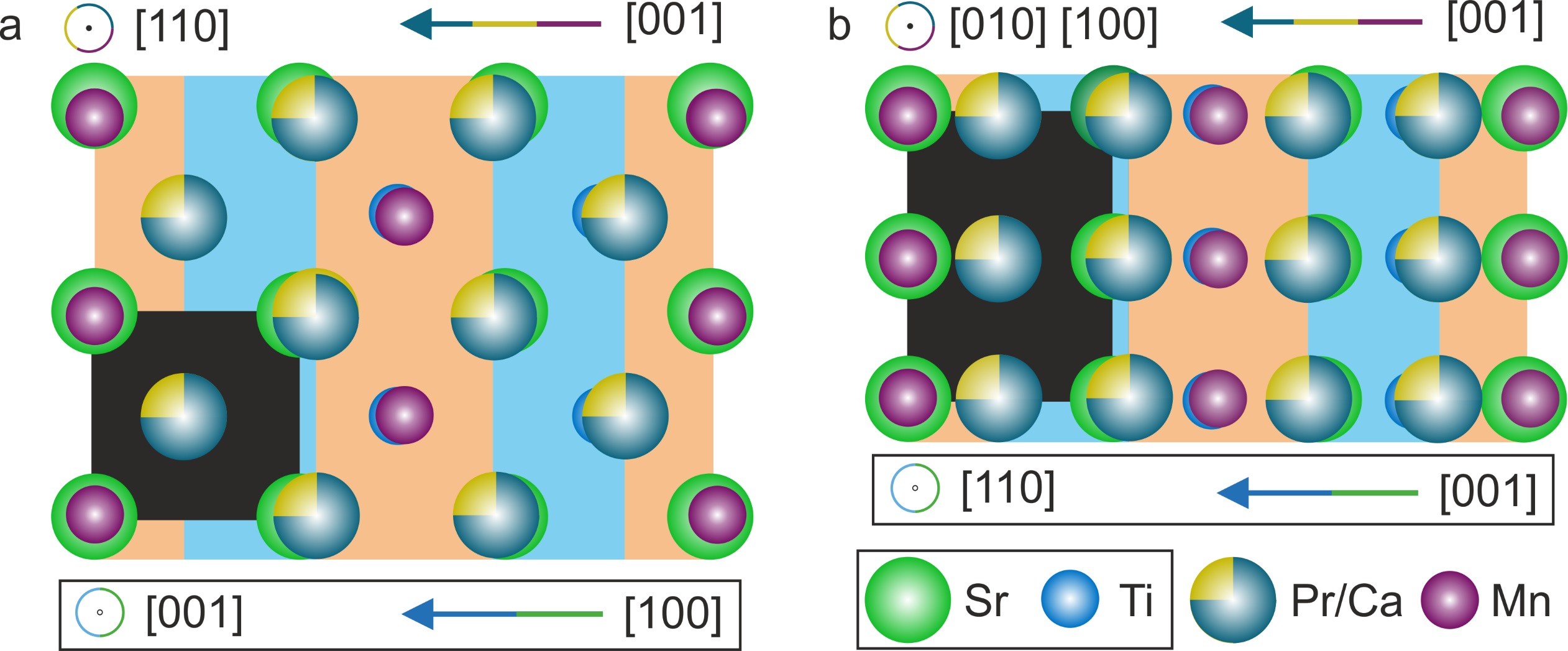}
			\caption{Schematic of the epitaxial relation for both substrates. The lattices are aligned regarding the lower left corner atom. The in-plane and out-of-plane directions of the RP phase and substrate are depicted above and beneath the drawing, respectively. The corresponding structural units are marked by colored boxes, where blue indicates rock-salt layers, orange perovskite layers and black STO unit cells. Note that the oxygen atoms are not depicted for reasons of clarity and that the sizes of the atoms are not to scale. (a) PCMO-RP film on a STO [001] substrate. The RP unit cell covers three unit cells in STO [100]-direction and two in [010]-direction. As apparent from the sketch in [100]-direction the film is exposed to compressive stress (1\%) and in [010] to tensile stress (-3\%). (b) PCMO-RP film on a STO [110] substrate. The RP unit cell covers three cells in [001]-direction and one in [100]/[010] direction. In [001]-direction the film is exposed to compressive stress (1\%) and in [100]/[010] to tensile stress (-3\%).}
			\label{fig:S_Lattice_matching}
			\end{figure}

\end{suppinfo}

%%%%%%%%%%%%%%%%%%%%%%%%%%%%%%%%%%%%%%%%%%%%%%%%%%%%%%%%%%%%%%%%%%%%%%
%%% The appropriate \bibliography command should be placed here.
%%% Notice that the class file automatically sets \bibliographystyle
%%% and also names the section correctly.
%%%%%%%%%%%%%%%%%%%%%%%%%%%%%%%%%%%%%%%%%%%%%%%%%%%%%%%%%%%%%%%%%%%%%%

\end{document}